\documentstyle[12pt,fleqn]{article}
\font\fonta=cmr12 scaled\magstep2

\font\fontc=cmr12 scaled\magstep1

\def\beq {\begin{eqnarray}}
\def\eeq {\end{eqnarray}}
\def\beqn {\begin{eqnarray*}}
\def\eeqn {\end{eqnarray*}}
\def\neqn {\nonumber}

\def\ni {\noindent}

\def\PL #1 #2 #3 {Phys. Lett.~{\bf#1} (#2) #3}
\def\NP #1 #2 #3 {Nucl. Phys.~{\bf#1} (#2) #3}
\def\ZP #1 #2 #3 {Z.~Phys.~{\bf#1} (#2) #3}
\def\PR #1 #2 #3 {Phys. Rev.~{\bf#1} (#2) #3}
\def\PP #1 #2 #3 {Phys. Rep.~{\bf#1} (#2) #3}
\def\PRL #1 #2 #3 {Phys. Rev.~Lett.~{\bf#1} (#2) #3}
\def\etal {{\it et al}.}

\def\ie {{\it i.e}.}
\def\qcondensate{$<{\bar q}q>$}

\def\qqbar{$q{\bar q}$}
\def\ccbar{$c{\bar c}$}
\def\bbbar{$b{\bar b}$}
\def\eps  {$\epsilon$}
\def\GeV{\mbox{GeV}}
\def\MeV{\mbox{MeV}}
\def\fm{\mbox{fm}}
\begin{document}

\begin{titlepage}

\def\boxit#1{\vbox{\hrule\hbox{\vrule\kern3pt
\vbox{\kern3pt#1\kern3pt}\kern3pt\vrule}\hrule}}
\setbox1=\vbox{\hsize 26pc  
\centerline{\bf {\large  ${\cal T}$}okyo {\large ${\cal M}$}etropolitan 
{\large ${\cal U}$}niversity}
\centerline{ \bf \small Nuclear Theory Group, Department of Physics} 
}
$$\boxit{\boxit{\box1}}$$
\vskip0.3truecm
\hrule width 16.0 cm  \vskip1pt \hrule width 16.0 cm height1pt
\vskip3pt

\baselineskip=0.4cm
\begin{flushright}{TMU-NT940803\\hep-ph/9408338\\August 1994}
\end{flushright}
\vspace {0.5cm}
%
%
%%%%%%%%%%%%TITLE%%%%%%%%%%%%%%
\baselineskip=0.9cm
\begin{center}{\fonta 
Heavy Quarkonium in \\the dual Ginzburg-Landau theory of QCD} 
\end{center}
%
%%%%%%%%%%%%NAME%%%%%%%%%%%%%
\vskip 1.0cm
\baselineskip=0.8cm
\begin{center}{\fontc 
Katsuhiko Suzuki\footnote{\ni e-mail 
address : ksuzuki@atlas.phys.metro-u.ac.jp} 
and Hiroshi Toki\footnote{\ni also RIKEN, Wako, Saitama 351, Japan}

\vskip 0.5cm
{\em Department of Physics, Tokyo Metropolitan University\\
Hachiohji, Tokyo 192-03, Japan}
}
\end{center}
\vskip 1.5cm
%%%%%%%%%%%%ABSTRACT%%%%%%%%%%%%%%%%
\centerline{\bf Abstract}
\vspace{0.3cm}
\noindent
We study the masses and the leptonic decay widths of heavy quarkonium in 
the dual Ginzburg-Landau (DGL) theory of QCD, in which the 
abelian monopole condensation plays 
an  essential role for color confinement.  The effect of color screening 
due to the light quark pair creation is introduced by the infrared 
momentum cutoff in the gluon propagator.  
We find that the color screening effect is important to reproduce 
the properties of the heavy quarkonium, in particular the leptonic 
decay width.  
We also discuss the underlying systematics of the heavy quark 
spectroscopy in the DGL theory.

\end{titlepage}

%%%%%%%%%%%%%%%%%%%%%%%%%%%%%%%%%%%%%%%%%%%%%%

\baselineskip=0.8cm

\ni
{\fontc {\bf 1 Introduction}}

It is our strong desire to understand the mechanism of color 
confinement and to construct a low energy effective theory of QCD.  
Without the knowledge of 
confinement, we can not describe hadrons and nuclei in terms of the 
fundamental quark degrees of freedom as well as can not predict the 
quark-gluon phase transition under extreme conditions.  
The abelian projection of QCD proposed by 't~Hooft is 
one of the most promising method to construct an effective theory 
of QCD with quark confinement\cite{tHooft}.  
In this approach, QCD reduces to an abelian 
$[U(1)]^2=U(1)_3^e \times U(1)_8^e \,$ theory with the gauge field 
singularity, which is identified as 
the magnetic monopole.  Recently, this idea is formulated 
by the Kanazawa group as a dual Ginzburg-Landau (DGL) theory of 
QCD\cite{Kanazawa}.  
By using the DGL theory, the quark-antiquark potential is derived 
within the static source approximation\cite{Kanazawa,Suganuma}, 
which may be realized in the heavy 
quarkonium.  The resulting potential shows the 
linear quark confinement when the QCD monopole 
condensation takes place.  
In addition, lattice QCD simulations in 
the maximal abelian gauge\cite{Review} indicate the condensation of 
the QCD monopole in the confinement phase\cite{Kron,Hioki1}, 
and the abelian dominance for physical quantities related with the 
confinement such as a string tension\cite{Hioki2}.  
These results support the 't~Hooft's conjecture and the validity 
of the DGL theory of QCD.

Beyond the quenched (static quark) approximation, roles of dynamical 
light quarks are important.  For the description of the light hadron 
structure, 
in which chiral symmetry and its  spontaneous breaking are essential, 
we must solve the Schwinger-Dyson (SD) equation and the 
Bethe-Salpeter equation with the $[U(1)]^2$ field equations together.  
Even if for the heavy quark system, the light quark degrees of 
freedom are crucial, 
since they produce {\it color screening} of the confinement force.    
Due to the spontaneous pair creation of light quarks in the flux 
tube, the confinement force between 
a heavy quark and antiquark pair is screened.  
As a result, heavy-light quark mesons are created, and the heavy 
quark potential 
should be flattened at the large distance longer than 
a typical hadronic size.  
In fact, such a color screening is observed in the lattice QCD 
calculations with the dynamical fermion, where the confinement 
potential shows a saturation for $r > 1 \, \fm$\cite{Lattice}.  
Recently, the color screening effect is incorporated within 
the DGL theory by introducing an infrared momentum cutoff in the 
non-perturbative gluon propagator\cite{Suganuma}.  
Here, we study how the color screening affects 
the physics of the heavy quarkonium using the DGL theory of QCD.

The spectroscopy of heavy quarkonia, {\ie}  {\ccbar} and {\bbbar}, 
is extensively studied within the quark potential 
models\cite{Models}.  
Such a phenomenological potential assumes the short range coulomb 
and the long range linear confinement force including the  
the color spin-spin, spin-orbit, and tensor interactions, 
and gives successful descriptions of heavy quarkonia.  
However, those studies reveal the following difficulties; 

\ni
1) The leptonic decay widths of {\ccbar} $S$-wave excited states, 
$\psi (3685)$, $\psi (4040), \; \cdots$, 
are systematically overestimated in any potential 
model\cite{Models,QPC}, though those 
models reproduce the width of 
the ground state $J/\psi$ as well as the 
bottomnium leptonic decay width.  (See Table 2 in ref. \cite{QPC}.)

\ni
2) Conventionally, $\psi$(4160) is assigned to the $^3 D_1$ state 
of the charm quark-antiquark bound state\cite{Models}, and therefore 
the $e^+ e^{-}$ annihilation width of this 
state is expected to be small, 
in comparison with the width of $S$-state.  
However, the experimental value 
is quite large, and is the same order of magnitude as those of the 
$S$-states.

In this paper, we demonstrate how these 
difficulties are overcome by the 
introduction of the color screening in the DGL theory.  
For simplicity, 
we ignore the spin-dependent interaction, 
and consider the spin averaged 
mass spectrum.  In order to include the spin and the angular 
momentum dependence, we must introduce the additional 
spin-dependent forces 
within the DGL theory as discussed in ref. \cite{DQCDpotential}.  

\vspace{1.5cm}

\ni
{\fontc {\bf 2 Heavy quark potential with color screening in DGL}}

We briefly show the heavy quark potential in the DGL theory, and the 
introduction of the color screening effect to the DGL potential.  
In the framework 
of the DGL theory, the infrared effective lagrangian of QCD is given 
by\cite{Kanazawa}, 
\vglue 0.05cm
\beq
{\cal L}_{DGL}  &=& \bar{\psi}[i\gamma^{\mu}(\partial_{\mu} +i
e\vec A_\mu \cdot \vec H) - m]\psi 
-{1 \over {2n^2}}[n\cdot (\partial \wedge \vec A)]^2
   -{1 \over {2n^2}}[n\cdot (\partial \wedge \vec B)]^2 \neqn \\
& &-{1 \over {2n^2}}[n\cdot (\partial \wedge \vec A)]^\lambda 
  [n\cdot ^*(\partial \wedge \vec B)]_\lambda 
  +{1 \over {2n^2}}[n\cdot (\partial \wedge \vec B)]^\lambda 
  [n\cdot ^*(\partial \wedge \vec A)]_\lambda \neqn \\
& & +\sum\limits_{\alpha =1}^3 
{[|(i\partial _\mu -g\vec \varepsilon _\alpha 
\cdot \vec B_\mu )\chi _\alpha |^2-\lambda 
(|\chi _\alpha |^2-v^2)^2]} \;\; ,
\label{lag}
\eeq
\vglue 0.1cm
\ni
where $\psi$ is the quark field with the current mass $m$, 
$\vec A _\mu= (A^3_\mu, A^8_\mu)$  the abelian gluon field 
with the color operator $\vec H = (\lambda^3/2, \lambda^8/2)$, and 
$\vec B _\mu= (B^3_\mu, B^8_\mu)$ the corresponding dual field.  
The monopole field $\chi _\alpha$ has a magnetic charge $g\vec 
\varepsilon_\alpha$ with the Dirac condition $eg = 4\pi$, 
and $n_\mu$ 
is arbitrary constant vector.  The QCD monopole self-interaction is 
phenomenologically introduced in the last term of (\ref{lag}), and 
parameters, $\lambda$ and $v$, are determined later.  
The DGL lagrangian (\ref{lag}) has $[U(1)^e]^2 \times 
[U(1)^m]^2$ symmetry, but $\vec A _\mu$ and $\vec B _\mu$ are not 
independent.  If the magnetic monopole condensation takes place 
in the QCD vacuum, {\ie} the 
magnetic $[U(1)^m]^2$ symmetry is spontaneously broken, 
the monopole and 
the dual gluon fields acquire their masses, 
$m_\chi = 2 {\sqrt \lambda} v$ and $m_B = {\sqrt 3} g v $.   
As a result, 
the color electric charge is confined by the dual Meissner effect.  
Integrating (\ref{lag}) over $\vec B _\mu$, one gets 
a diagonal (abelian) gluon propagator\cite{Kanazawa,Suganuma},
%
%eq. (1)
%
\vglue 0.1cm
\beq
D_{\mu \nu}(k) = -\frac{1}{k^2} \{g_{\mu \nu} +(\alpha_e -1) 
\frac{k_{\mu} k_{\nu}}{k^2} \}  + 
\frac{1}{k^2} \frac{m_{B}^2} {k^2 - m_{B}^2} \frac{n^2}
{(n \cdot k)^2 }   X_{\mu \nu}(k) \; \; ,
\label{gluon0}
\eeq
\vglue 0.1cm
\ni
where
\beqn
X^{\mu \nu }=-{1 \over {n^2}}\varepsilon _\lambda 
^{\mu \alpha \beta }\varepsilon ^{\lambda \nu \gamma \delta }
\;n_\alpha n_\gamma \,k_\beta k_\delta 
\eeqn
\vglue 0.2cm
\ni
Here, $m_B$ is the dual gluon mass due to the abelian monopole 
condensation.  In the static (heavy) 
quark-antiquark system,  we assume the vortex solution of 
the type-II 
superconductor.  Thus, (\ref{gluon0}) leads to the 
potential as\cite{Kanazawa,Suganuma},
%
%eq. (2)
%
\vglue 0.1cm
\beq
V(r)&=& V_{\mbox{Yukawa}}(r) + V_{\mbox{Linear}}(r) \\
 & &\hspace{0.5cm}  V_{\mbox{Yukawa}}(r) =
- \frac{\vec Q^2} {4\pi} \frac{e^{-m_{B} r}} {r}\\
& &\hspace{0.5cm}  V_{\mbox{Linear}}(r) = 
\frac{\vec Q^{2} m_{B}^2} {8\pi}
     \,  \mbox{ln}(\frac{m_{B}^2 + m_{\chi}^2} {m_{B}^2}) \; r, 
\eeq
\vglue 0.1cm

\ni
where $\vec Q^{2}$ is the sum of the diagonal color charge, 
$\vec Q^{2} = Q_{3}^{2} + Q_{8}^{2} = {e^2}/{3}$.  
If the magnetic $U(1)_{3}^m \times U(1)_{8}^m$ symmetry 
were not broken, 
{\ie} dual gluon mass $m_{B}=0$, the potential would be a pure 
coulomb type.  
The empirical value 
of the string tension $\sim$1GeV/fm is reproduced with 
$m_{B} \sim 0.5 \GeV$ and $m_{\chi} \sim 1 \GeV$.

Up to now, we work with the quenched approximation, and do not deal 
with an effect of the color screening.  
In the presence of the dynamical light quarks, the string of the 
hadron flux 
becomes unstable against the  spontaneous {\qqbar}  pair creation.  
This means that the flux tube is divided into two pieces when the 
distance between the 
valence quark and the antiquark becomes larger than a certain 
critical value $R_{SC} \sim 
1 \fm$.  Indeed, the results of the lattice QCD simulation with the 
dynamical light quarks show that 
the {\qqbar}  potential becomes flattened for 
$r > 1\fm$\cite{Lattice}.

In the DGL theory, we introduce the color screening effect by the 
following simple replacement in the gluon propagator, as done in 
ref. \cite{Suganuma};
\vglue 0.1cm
\beq
D_{\mu \nu}(k) = -\frac{1}{k^2} \{g_{\mu \nu} +(\alpha_e -1) 
\frac{k_{\mu} k_{\nu}}{k^2} \}  + \frac{1}{k^2} 
\frac{m_{B}^2} {k^2 - m_{B}^2} \frac{n^2}
{(n \cdot k)^2 + \epsilon ^2}   X_{\mu \nu}(k)
\label{gluon1}
\eeq
%
%
%
%\vglue 0.4cm

\ni
Here, the parameter $\epsilon$ plays an infrared momentum cutoff 
in the gluon propagator, and removes the double pole structure 
$1 \over {{(n \cdot k)}^2}$, which 
plays an essential role for the long 
range quark correlation.  
In principle, this infrared cutoff $\epsilon$ 
can be obtained by calculating the quark polarization diagram after 
getting the non-perturbative quark propagator, 
though it is hard to handle it.  
In ref. \cite{Suganuma}, the energy of the created {\qqbar} pair  
was estimated $<2 E_{q}>$ $\sim$ 850 MeV 
by using the Schwinger mechanism.  
Since the energy of created pair is 
supplied by cutting off the hadronic string,  
the critical distance $R_{SC}$ 
may be represented as $k \; R_{SC} \; \sim \; <2 E_{q}>$, 
where $k$ is the string 
tension $\sim$ 1GeV/fm.  
Then, we may take the critical distance of order of 
$R_{SC}$ $\sim$  ${1/\epsilon}$ $\sim$ 1fm.
In the present study, we treat the infrared cut $\epsilon$ as a free 
parameter of the model; {\eps} $\sim$ 100MeV.  
Using (\ref{gluon1}), we get the modified heavy quark potential with 
the color screening effect.  
\vglue 0.1cm
\beq
V(r)^{SC} = - \frac{\vec Q^2} {4\pi} \frac{e^{-m_{B} r}} {r} \, + \,
\frac{\vec Q^{2} m_{B}^2} {8\pi}
                  \frac{1-e^{-\epsilon r} }{\epsilon} \, \mbox{ln}
(\frac{m_{B}^2 + m_{\chi}^2 - \epsilon^2} 
{m_{B}^2- \epsilon^2} ) \; , 
\eeq
\vglue 0.4cm

\ni
The resulting quark potential for $\epsilon$=0, 100, and 200 MeV is 
shown in Fig.1.  
The short range Yukawa part is not modified by the color screening.  
Note that, in the long distance limit; $r \gg 1/ \epsilon$, the quark 
potential tends to be constant, 
and shows a saturation behavior which 
agrees with the lattice QCD calculation.  
\vglue 0.1cm
\beq
V(r)^{SC}_{\mbox{Linear}} \rightarrow 
\frac{\vec Q^{2} m_{B}^2} {8\pi} \;
                  \frac{1}{\epsilon} \; \mbox{ln}
(\frac{m_{B}^2 + m_{\chi}^2 - \epsilon^2} {m_{B}^2- \epsilon^2} ) \;
= \mbox{constant}
\eeq
\vglue 0.4cm

As shown above, the screening of the confinement force is naturally 
incorporated in the DGL theory of QCD.   
We remark here that the role of 
the infrared cutoff {\eps} is crucial 
to solve the SD equation for the 
light quarks, since it removes the infrared double pole structure in 
the gluon propagator\cite{Suganuma}.   We next consider the heavy 
quarkonium spectroscopy in the DGL theory, 
and study the effect of the 
color screening on the heavy quarkonium.

Our present study shares some similarities with the recent work of 
Yu-bing {\etal}\cite{Screen}, 
where the color screening effect for the charmonium is studied in 
terms of the error function type confinement force.  
However, our treatment is fully based on the DGL theory, which 
has a clear link with QCD by virtue of the abelian projection 
procedure.  
The infrared cutoff parameter $\epsilon$ is not merely a 
phenomenological 
one, but calculable in the DGL theory.  
Moreover, we can also study the light hadron physics 
in terms of the gluon propagator (\ref{gluon1}), 
which enables us to construct a consistent description of 
light and heavy quark systems.  In fact, we will show later that the 
infrared cut $\epsilon$, which is fixed 
by the masses and the leptonic decay widths of the heavy quarkonium, 
also reproduces correct values of light hadron properties such as 
the pion decay constant $f_{\pi}$ and the quark 
condensate \qcondensate.

\vspace{1.5cm}

\ni
{\fontc {\bf 3 Mass spectrum and leptonic decay width}}

We calculate the mass spectra of the heavy quarkonium in terms of 
the quark potential obtained in the previous section.  
In order to get the 
energy spectrum of bound states, 
we use the diagonalization procedure by 
using the harmonic oscillator wave function.   
We neglect the spin dependence of the 
spectrum for simplicity, and discuss the spin averaged spectrum, 
since the energy shift due to the spin dependent interaction is 
of the order of 10 MeV due to the large quark mass.    First, 
we show in Fig.2 the color screening dependence of the 
charmonium spectrum.  Here, we use the parameters 
$m_B = $ 500 MeV, $m_{\chi}$ = 1458 MeV,  $e$ = 5.45, 
the charm quark mass 1289 MeV, and the infrared cutoff 
$\epsilon =$ 0, 100, and 200 MeV.  
It is clear that the $1S$ ground state (corresponds 
to $J/\psi$ and $\eta_c$) 
is almost independent of the choice of $\epsilon$, whereas the 
spectrum of $S$-wave excited states and 
$P$-, $D$-states receive substantial 
contributions.  Since the change of 
the confinement potential due to the color 
screening is appreciable for $r > 1 \fm$, 
the 1$S$ state is not affected 
due to its small radius.  The 2$S$, 1$P$, and 1$D$  become loosely 
bound by the reduction of the confinement force, 
and then the energies of these 
states decrease.  To see this behavior clearly, we plot in 
Fig.3 the wave functions of $1S$, 2$S$, 1$P$, and 1$D$ states with 
$\epsilon$ = 0, 100, and 200 MeV.  Apparently, the 1$S$ wave 
function is unchanged as expected.  Other states show 
considerable $\epsilon$-dependence, in particular, the value at 
the origin $|\Psi(0)|$ decreases as $\epsilon$ increases.  
The similar  behavior is also found in Fig.4, 
where the {\eps}-dependence of the root mean square radii and 
$\frac{|\Psi_{nS}(0)|^2}{|\Psi_{1S}(0)|^2}$ are shown.  
The root mean square radii are normalized to their 
{\eps} = 0 values.  The radii of those states at 
{\eps} = 0 MeV are $<r^2>^{1/2}_{1S}$ = 0.46 fm, 
$<r^2>^{1/2}_{1P}$ = 0.72 fm, 
$<r^2>^{1/2}_{2S}$ = 0.90 fm, and $<r^2>^{1/2}_{1D}$ = 0.93 fm.
The color screening effect makes the sizes of bound states large, 
and decreases the 
values at the origin of the $S$-wave excited states.  
This behavior is of 
great significance for the discussion of  the
 $e^+ e^- $ annihilation widths.

In order to study the color screening effect in detail, 
we compare two cases; $\epsilon$ = 0 and $\epsilon$ = 100 MeV.  
For the model parameters, 
we fix the masses of the dual gluon and the 
magnetic monopole, the gauge coupling $e$, and the charm quark mass  
to reproducing the masses of $1S$, $1P$, $2S$, and 
1$D$ of the charmonium; 
$m_B$=500(516)MeV, $m_\chi$=1524(1704)MeV, $e$=4.8(5.2), and 
$m_c$= 1278(1253)MeV for {\eps}=0(100)MeV.  
In both cases, the mass spectrum is well reproduced, 
as tabulated in Table 1.  
Note that $\psi(4160)$ is assigned to 4$S$-state 
in the {\eps} = 100 MeV case, which 
is different form the conventional 2$D$ assignment.  
The masses of $S$-wave excited states decrease 
due to the screening of the linear confinement 
potential so that above assignment becomes possible.

The $e^+ e^- $ annihilation decay width is given by the 
following formula including the QCD correction\cite{Decay};
\vglue 0.1cm
\beq
\Gamma_{nS}(1^{-} \rightarrow e^{+} e^{-} ) = \frac{16 \pi e_{i}^2 
\alpha^2} {M_{nS}^2} |\Psi_{nS} (0)|^2 
(1-\frac {16 \alpha_{S}}{3 \pi}) \;\; .
\label{decay}
\eeq
\vglue 0.4cm

\ni
To avoid the  ambiguity of the  choice of the QCD correction, 
we concentrate on 
the ratio of $nS$ to 1$S$ decay width.  This ratio corresponds to 
\vglue 0.1cm
\beq
\frac {\Gamma_{nS}} { \Gamma_{1S}} = (\frac{M_{1S}}{M_{nS}})^2 
\frac{|\Psi_{nS}(0)|^2}
{|\Psi_{1S}(0)|^2}  \;\; .
\eeq
\vglue 0.4cm

\ni
From the comparison with experiment, 
most of the theoretical calculations 
overestimates the decay widths for 3$S$, 4$S$ states \cite{Models}.  
Our model calculations with {\eps} = 0 MeV gives also larger values 
than the experimental data as found in the third column of Table 2.  
However, in the case {\eps} = 100 MeV, the decay width ratio is 
fairly reproduced as shown in the fourth column, 
since the color screening effect reduces the values of $|\Psi(0)|$ 
as already depicted in Fig.3.  
Particularly, the results of two cases for $\psi(4160)$ show a 
striking difference.  In the case of {\eps} = 100 MeV, 
the calculated width 
is in a good agreement with data due to the 4$S$-state assignment of 
$\psi(4160)$.   On the other hand, the leptonic decay width vanishes 
in the {\eps} = 0 MeV  case, since the value of 
the 2$D$ wave function at the origin is zero.  Of course, inclusion 
of the tensor interaction in the potential  
causes the $S$-$D$ mixing of the wave function, and 
hence  the width of the {\eps} = 0 MeV case becomes non-zero.  
However, such a $S$-$D$ mixing effect  gives a still much smaller 
width than experimental data (about 10 \% at most) 
within the realistic tensor force\cite{QPC}.

It is quite interesting to note that the use of 
$\epsilon \sim $100 MeV also leads 
good results for light quark properties such as the pion decay 
constant $f_{\pi} \sim$ 80 MeV and the 
quark condensate $<{\bar q}q>_{RGI}
\sim (-200 \MeV )^3$, which are obtained 
by solving the Schwinger-Dyson equation\cite{Suganuma,Sasaki}.  
Both light and 
heavy quark properties are well reproduced 
in the DGL theory with the color screening.  
This fact indicates a close connection between the confinement 
mechanism and the chiral dynamics of the light quark physics.

We finally address a systematics of 
the heavy quark spectroscopy using 
several type of the potential.  In Fig.5, we show the charmonium 
mass spectra in terms of the linear, the linear plus 
coulomb, and the DGL potential with the experimental data.  
The linear potential provides already 
quite a good description of the 
experimental spectrum.  The coulomb attraction is, however, 
needed to bring 
down especially the 1$S$ state relative to 1$P$ state.  
The higher $S$ states 
are overestimated with the linear plus coulomb potential.  
We note that $\psi(4160)$ was identified to the 2$D$ state 
due to this agreement with 
calculation\cite{Models}.  
If we adopt now the DGL potential with the color 
screening, the higher $S$ states could be brought down largely.  
This fact makes another assignment of $\psi(4160)$ 
to the 4$S$ state\cite{QR}.  If this is 
the case, we can solve the puzzle of the large leptonic 
decay width of $\psi(4160)$ as already done in Table 2.

Here, we comment on the properties of {\bbbar} in the DGL theory.  
We get reasonable agreements with experimental data using the bottom 
quark mass $m_b =$ 4734 MeV for {\eps}=0 MeV.  
The calculated root mean square radii are 
$<r^2>^{1/2}_{1S}$ = 0.20 fm and  $<r^2>^{1/2}_{1P}$ = 0.41 fm 
for {\eps} = 0 MeV, which are much smaller than those of 
the charmonium.  Since the DGL theory is designed as 
the infrared effective theory of QCD, 
it is not adequate to discuss such a 
short range structure $<r^2>^{1/2} \sim$ 0.2 fm $\sim$ 1GeV$^{-1}$.  
Clearly, the effect of the color screening is not so important in 
the bottomnium case due to those small radii.  
Therefore, we do not discuss the bottomnium spectroscopy here 
in terms of the DGL theory.  
In order to investigate the {\bbbar}, we must consider the short 
range gluon exchange, {\ie} coulomb type interaction, in which the 
role of non-diagonal gluon is crucial.  
We have to develop a theory, 
which connects smoothly the long distance physics 
discussed here and the short range perturbative physics.  
Such a study is beyond the scope of the present work.

\vspace{1.5cm}

\ni
{\fontc {\bf 4 Summary and discussions}}

We have studied the heavy quarkonium properties in the dual 
Ginzburg-Landau theory of QCD, 
which generates the color confinement due 
to the abelian monopole condensation.  The color screening of the 
confinement potential due to the light quark pair creation has been 
incorporated 
by introducing an infrared momentum cutoff in the abelian gluon 
propagator.  Such an infrared momentum cut controls the 
quark long range correlation.

We have found that the DGL theory gives a good description of heavy 
quarkonia with $\epsilon \sim $100 MeV.  
The screening of the confinement 
force reduces the leptonic decay width of the $nS(n>2)$-states.  
This potential brings down also the higher $S$ states, 
and make it possible 
to identify $\psi(4160)$ as the 4$S$ state instead of 2$D$ state.  
This identification solve the problem of 
the large $e^+ e^-$ decay width of $\psi(4160)$.

Further studies on the $ {\bar D} D$ decay of the charmonium and 
$ {\bar B} B$ decay of the bottomnium are of special interest within 
the DGL theory.  
For those processes, the quark pair 
creation model\cite{QPC,QPC2} can reproduce the experimental data.  
In such a study, however, a probability of the light quark pair 
creation from the vacuum is treated as a {\it free} parameter.  
In the DGL theory, the rate of the 
light quark pair creation can be estimated by using the Schwinger 
mechanism\cite{Suganuma}.  Realistic calculations of the pair 
creation  process are under investigation.  

It is very interesting and 
important to note that the choice of $\epsilon \sim $100 MeV also 
reproduces the pion decay constant and the quark condensate 
in the light quark physics\cite{Suganuma}.  
Commonly, the low energy hadron physics is believed to be 
dominated by the 
chiral dynamics, whereas the heavy quark system is governed by the 
linear confinement potential with the perturbative coulomb part, 
in which the chiral symmetry is less important.  
Hence, we have no clear connection between 
the confinement and the physics of the chiral symmetry.  
On the other hand, the lattice QCD at finite 
temperature tells us that the deconfinement transition 
and the chiral transition take place at the same temperature, 
and suggests that the quark confinement and the chiral dynamics 
should be strongly correlated.  
In the DGL theory of QCD, the quarks are confined by the dual 
Meissner effect due to the abelian monopole condensation.  
Here, we have shown that the heavy 
quark spectroscopy as well as the light hadron properties are 
reproduced by the {\it same} fundamental parameter of the DGL, 
{\ie} the monopole mass, the dual gluon mass, the gauge coupling, 
and the infrared momentum cut $\epsilon$.  
Such agreements support the DGL theory as 
the effective theory of QCD, which provides an unified treatment of 
the confinement and the dynamical chiral symmetry breaking.  
These findings motivate us 
to do further investigations of various low 
energy phenomena in terms of the DGL theory of QCD.

%%% TABLE CAPTIONS %%%
\newpage

\baselineskip=0.8cm

{\bf \large Table Captions}
\vglue 0.4cm

\ni
{\bf Table 1} : The charmonium mass spectrum in the DGL theory 
with color screening.  The 
experimental data\cite{Data} with the corresponding states are 
shown in the first column. 
Theoretical results with {\eps} = 0 MeV and 100 MeV 
are depicted in the second and the third columns, respectively.  
Note that we do not consider the spin-dependent interaction, 
and calculated values are the spin averaged ones.  

\vglue 1cm 

\ni
{\bf Table 2} : The ratio of $nS$ to 1$S$ leptonic decay width of 
the charmonium 
in the DGL theory.  The experimental data\cite{Data} are shown 
in the second column.  The results  with {\eps} = 0 MeV and 100 MeV 
are depicted in the third and the fourth columns, respectively.

%%% TABLES %%%
\newpage

{{\bf Table 1}}

\begin{tabular}{lcc}
& & \\ \hline 
State          & $\epsilon=0$      & $\epsilon = 100$  \\ \hline
$J/\psi $  & $1S$(3074)      & $1S$(3066)\\ 
$\chi_{cJ}(\sim 3500)$    & $1P$(3492)      & $1P$(3507)\\ 
$\psi (3685)$    & $2S$(3664)      & $2S$(3644)\\ 
$\psi (3770)$    & $1D$(3781)      & $1D$(3775)\\ 
$\psi (4040)$    & $3S$(4108)      & $3S$(3999)\\ 
$\psi (4160)$    & $2D$(4190)      & 4$S$(4256)\\   
$\psi (4415)$    & $4S$(4492)      & $5S$(4469)\\ \hline
\end{tabular}

\vglue 2cm

{{\bf Table 2}}

\begin{tabular}{lccc}
 & & &\\ \hline 
State  & Exp   & $\epsilon=0$ & $\epsilon = 100 \MeV$  \\ \hline
$\Gamma_{\psi (3685)} / \Gamma_{J/\psi}$   & $0.40 \pm 0.04$ & 0.48  
& 0.38  \\ 
$\Gamma_{\psi (4040)} / \Gamma_{J/\psi}$   & $0.14 \pm 0.03$ & 0.34 
& 0.22  \\ 
$\Gamma_{\psi (4160)} / \Gamma_{J/\psi}$   & $ 0.14\pm 0.04$ &
 0.0$^a$  & 0.18$^b$  \\ 
$\Gamma_{\psi (4415)} / \Gamma_{J/\psi}$   & $0.09 \pm 0.02$ & 0.27 
& 0.08  \\ 
\hline
\end{tabular}

\vglue 0.5cm
\ni
a) In this case $\psi(4160)$ is assigned to the 2$D$ state 
due to the mass spectrum.

\ni
b) In this case $\psi(4160)$ is assigned to the 4$S$ state 
due to the mass spectrum.

%
%%% FIGURE CAPTIONS %%%
\newpage

\baselineskip=0.8cm

{\bf {\large Figure Captions}}
\vglue 0.4cm

\ni
{\bf Fig.1} : The heavy quark potential in the dual 
Ginzburg-Landau theory with the color screening.  
The solid curve represents the case of $\epsilon$ = 0 MeV 
(no screening).  The dash-dotted and the dashed curves denote 
the $\epsilon$ = 100 and 200 MeV cases, respectively.  

\vglue 1cm 

\ni
{\bf Fig.2} : The infrared cutoff $\epsilon$ 
dependence of the charmonium spectrum.  

\vglue 1cm

\ni
{\bf Fig.3} : The wave functions of the {\ccbar} 
bound states for 1$S$, 2$S$, 1$P$, and 1$D$ 
states calculated in the DGL theory with the color screening.  
The results with $\epsilon=$ 0, 100, and 200 MeV are shown by 
solid, dash-dotted, and dashed curves, respectively.
\vglue 1cm

\ni
{\bf Fig.4} : The infrared cutoff {\eps} dependence of the root mean 
square radii and the values of the wave functions at the origin.  
The root mean square radii 
normalized to the {\eps}=0 MeV value, are shown by the dashed curves 
for $1S$, $1P$, $1D$, and $2S$ states.  
$\frac{|\Psi_{nS}(0)|^2}{|\Psi_{1S}(0)|^2}$ 
is also shown by the solid curves for the $2S$ and $3S$ states.  
%for {\eps}=0, 100, and 200 MeV cases, 

\vglue 1cm

\ni
{\bf Fig.5} : The charmonium mass spectra in 
various potential models; from left to right the linear potential, 
the linear plus coulomb potential,  
the DGL theory with color screening, and experimental spectrum.  
In the experimental spectrum, the state at 4160MeV is 
either assigned to the 2$D$ state (a) and 4$S$ state (b).


\newpage
\baselineskip=0.8cm
\begin{thebibliography}{99}
%
\bibitem{tHooft}G. 't~Hooft, \NP B190 1981 455
%
\bibitem{Kanazawa}
T. Suzuki, Prog. Theor. Phys. {\bf 80} (1988) 929; 
{\bf 80} (1989) 752\\
S. Maedan and T. Suzuki, Prog. Theor. Phys. {\bf 81} (1989) 229
%
\bibitem{Suganuma}H. Suganuma, S. Sasaki and H. Toki, 
RIKEN AF-NP-164 (1994), hep-ph/9312350\\
H. Toki, H. Suganuma and S. Sasaki, in Proceedings of International 
Symposium on Spin-Isospin Responses and Weak Processes 
in Hadrons and Nuclei, eds. T. Suzuki {\etal}, 
to appear in Nucl. Phys. {\bf A} (1994)
%
\bibitem{Review}For a review, T. Suzuki, Nucl. Phys. 
{\bf B} (Proc. Suppl.) {\bf 30} (1993) 176, and references therein 
%
\bibitem{Kron}A.S. Kronfeld \etal, \PL B198 1987 516\\
A.S. Kronfeld, G. Schierholz and U.-J. Weise, \NP B293 1987 261
%
\bibitem{Hioki1}S. Hioki \etal, \PL B285 1992 343
%
\bibitem{Hioki2}S. Hioki \etal, \PL B272 1991 326
%
\bibitem{Lattice}E. Laermann \etal, \PL B173 1986 437\\
R. Gupta \etal, \PR D44 1991 3272\\
W. Sakuler \etal, \PL B276 1992 155
%
\bibitem{Models}E. Eichten \etal, \PR D21 1980 203\\
S. Godfrey and N. Isgur, \PR D32 1985 189 \\
For a review, {\it Quarkonia}, ed. W. Buchm{\"u}ller, North-Holland\\
For a comparison of models, see D. Besson and T. Skwarnicki, 
preprint CLNS 93/1239 (1994), and references therein\\
%M. Baker, J.S. Ball and F. Zachariasen, \PR D45 1992 910
%
\bibitem{QPC}K. Heikkila, N.A. T{\"o}rnvist, S. Ono, \PR D29 1984 110
%
\bibitem{DQCDpotential}M. Baker, J.S. Ball and F. Zachariasen, 
\PR 44 1991 3949,  {\it ibid} {\bf D47} (1993) 3021;  
\PL B283 1992 360\\
For a review, M. Baker, J.S. Ball and F. Zachariasen, preprint CALT 
68-1887 (1994)
%
\bibitem{Screen}Dong Yu-bing \etal, \PR D49 1994 1642
%
\bibitem{Decay}R. Van Royen and V. Weisskopf, Nuovo Cimento {\bf A50} 
(1967) 610\\
R. Barbieri \etal, \PL B57 1975 455
%
\bibitem{Data}Particle Data Group, K. Hikasa \etal, \PR D45 1992 S1
%
\bibitem{Sasaki}S. Sasaki, H. Suganuma and H. Toki, 
TMU preprint (1994)
%
\bibitem{QR}C. Quigg and J. Rosner, Phys. Rep. {\bf 56} (1979) 167 
%
\bibitem{QPC2}A. Le Yaouanc \etal, 
Hadron Transitions in the Quark model; 
Gordon and Breach (1988), and references therein
%
\end{thebibliography}
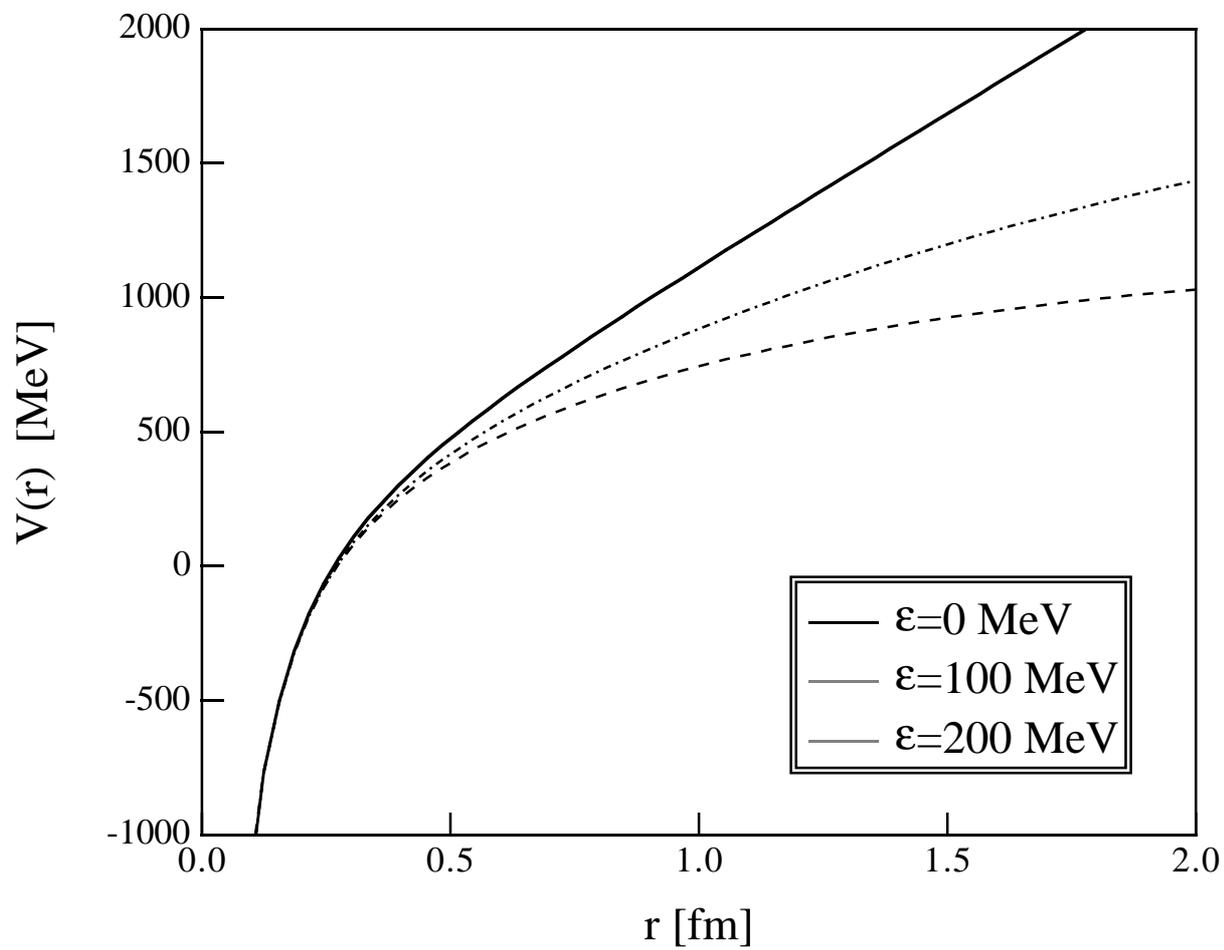
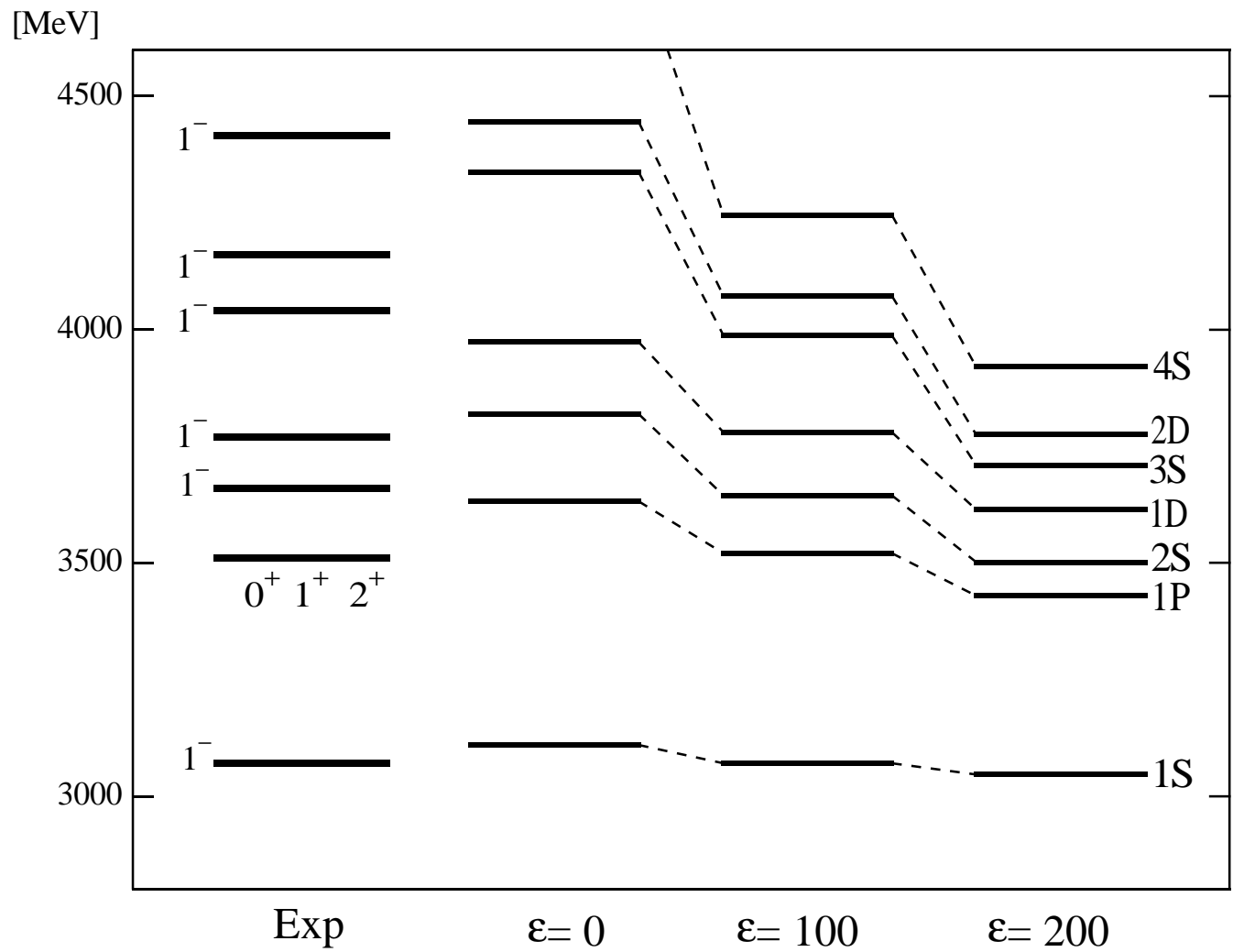
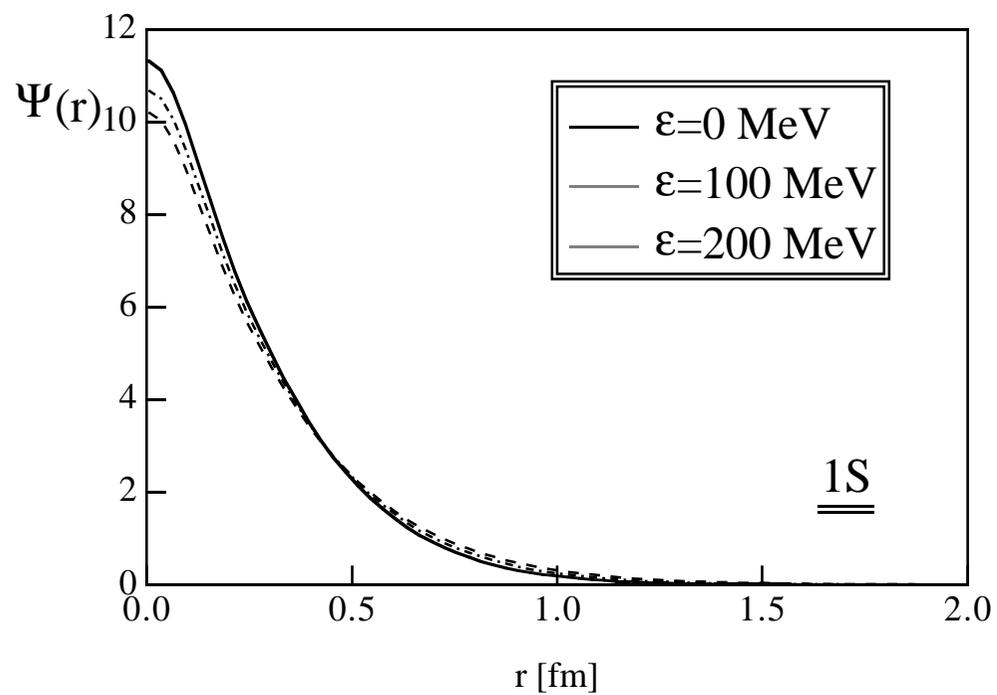
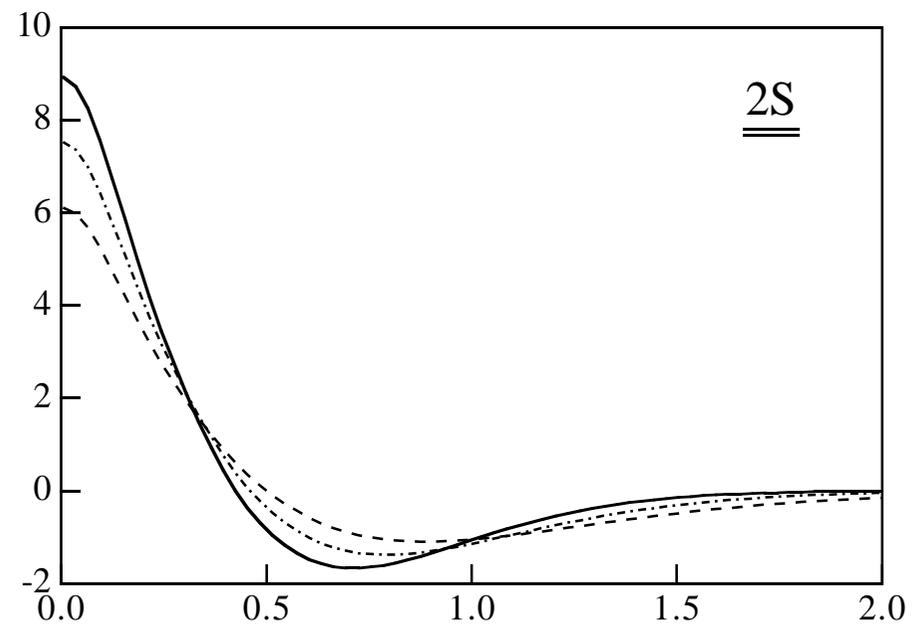
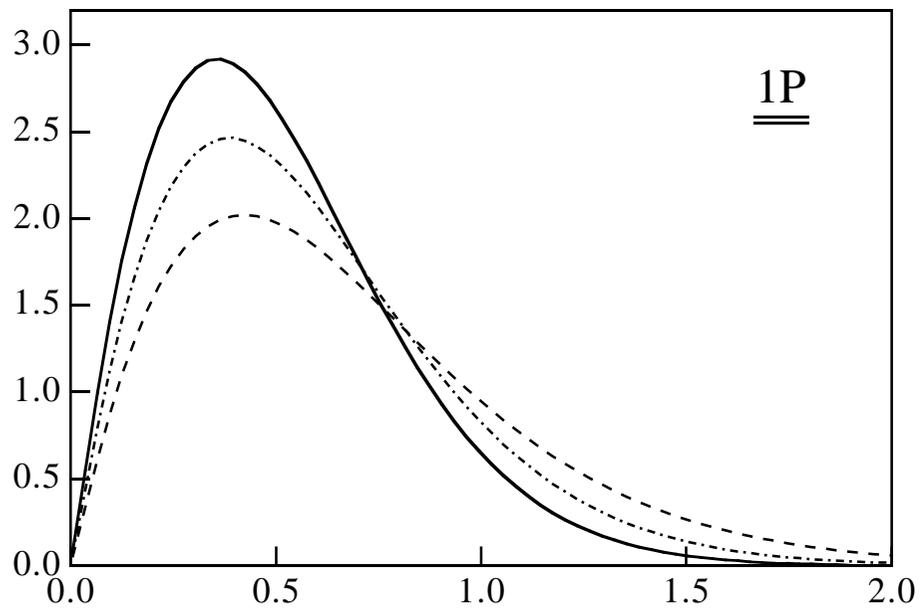
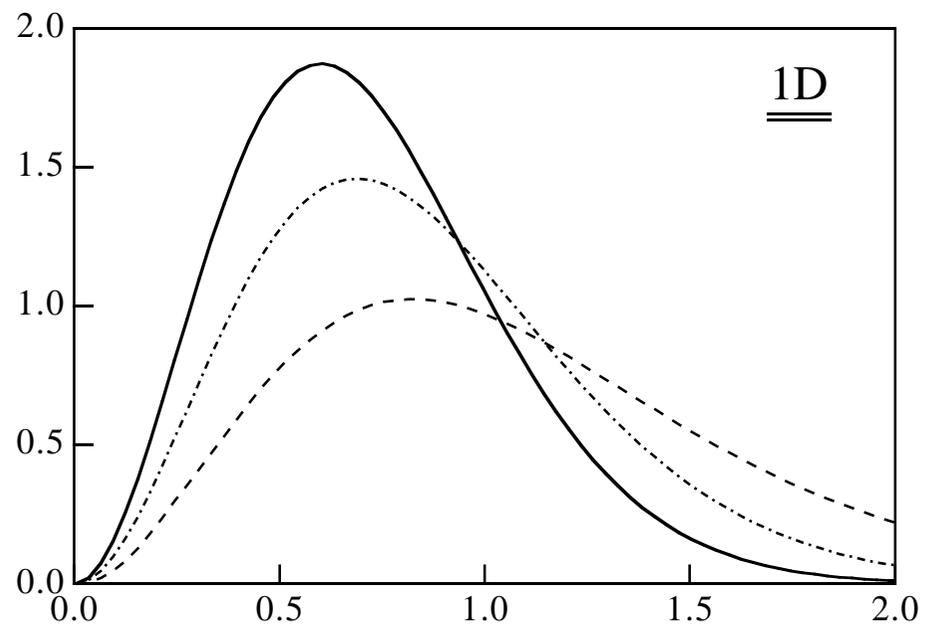
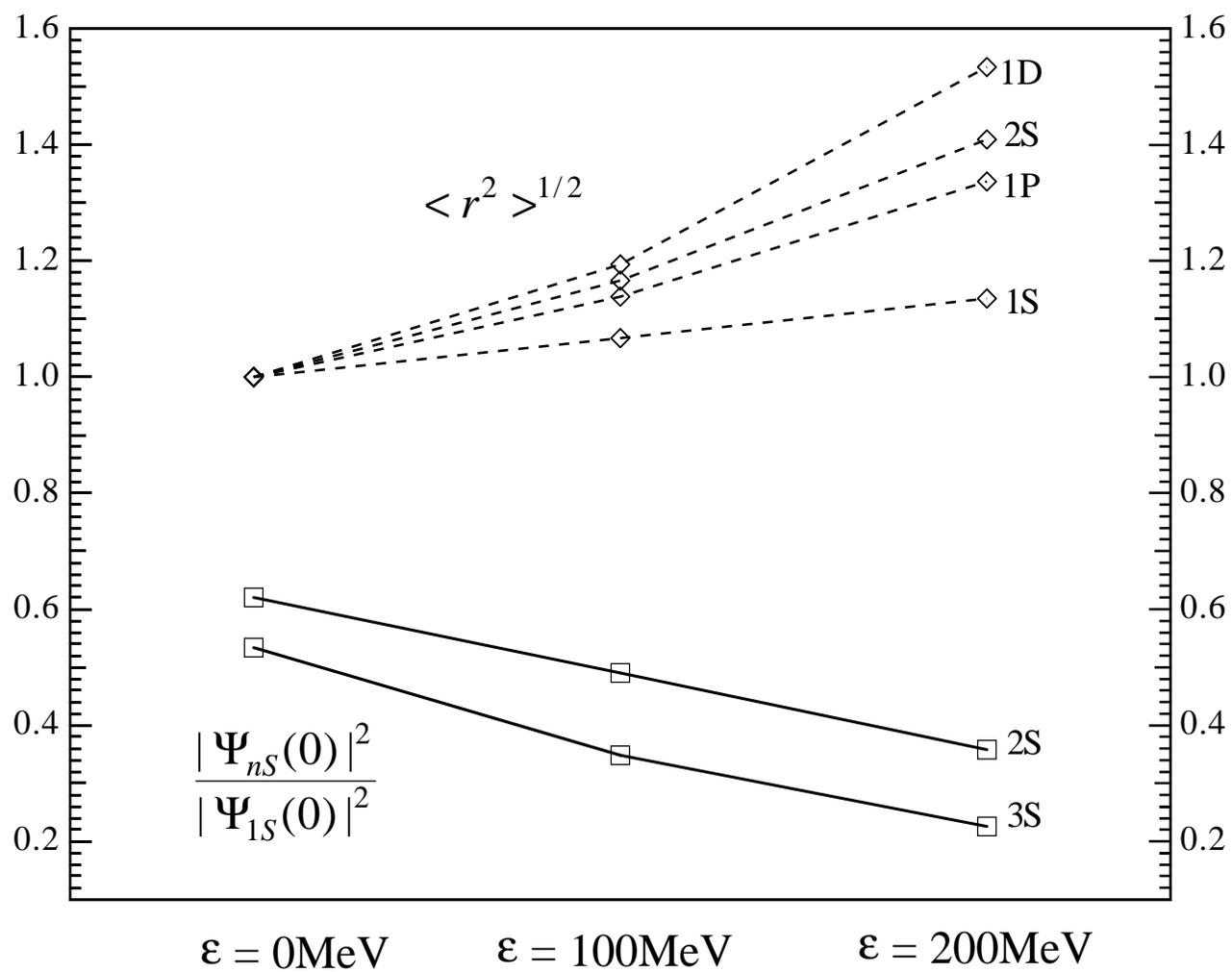
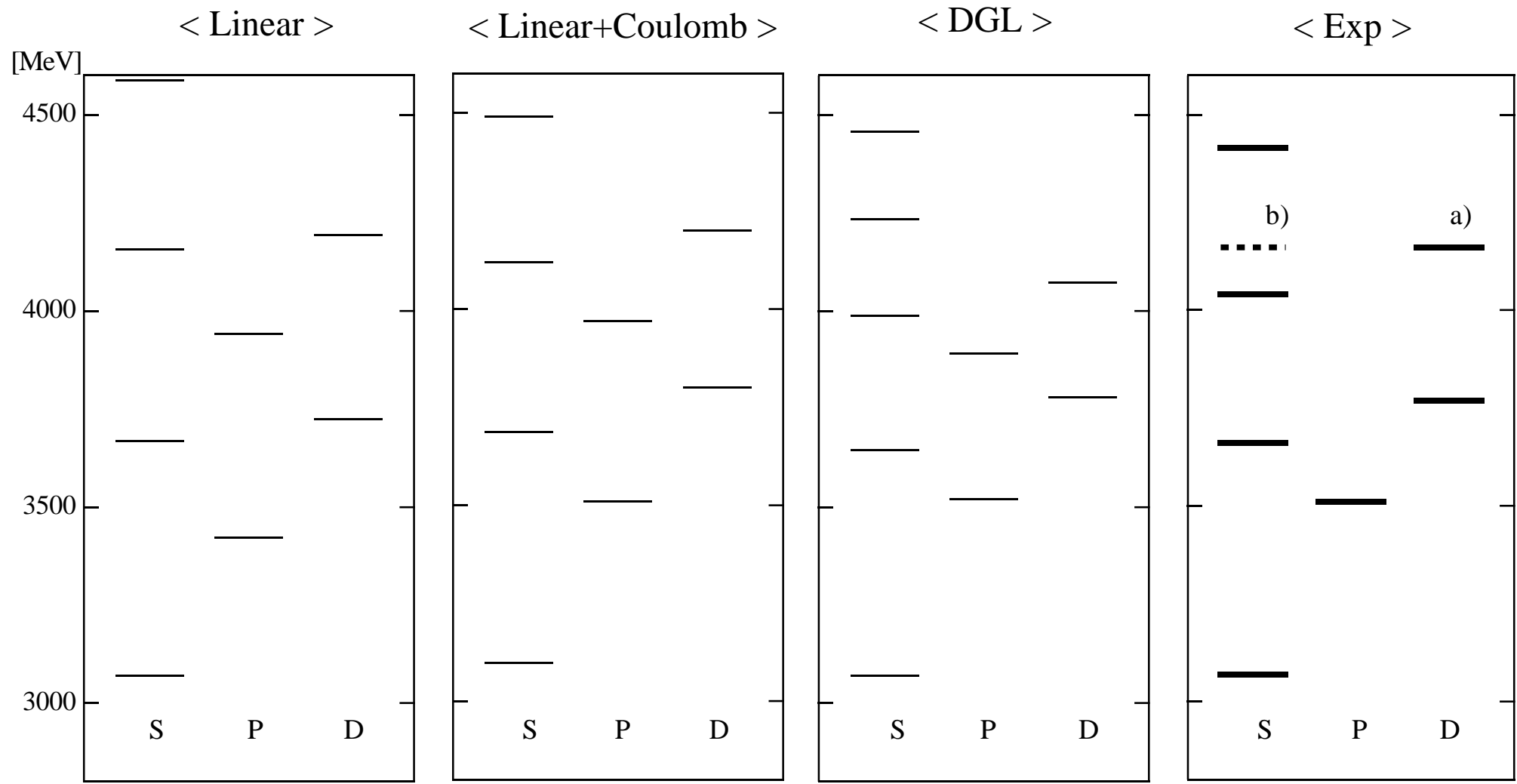
\end{document}